\documentclass[prd,showpacs,superscriptaddress,twocolumn,
floatfix,preprintnumbers,altaffilletter]{revtex4}

\usepackage{longtable}
\usepackage{graphicx}
\usepackage{amsmath,amssymb}
\usepackage{color}
\usepackage{units}
\usepackage{epstopdf}
\usepackage{hyperref}
\usepackage{multirow}
\usepackage{url}

\usepackage{subfigure}
\usepackage{rotating}

\newcommand{\beq}{\begin{equation}}
\newcommand{\eeq}{\end{equation}}
\newcommand{\bdm}{\begin{displaymath}}
\newcommand{\edm}{\end{displaymath}}

\definecolor{Gray}{gray}{0.9}
\definecolor{orange}{rgb}{0.9,0.5,0}

\graphicspath{{./plots/}}

\begin{document}

\title{On standardizing kilonovae and their use as standard candles to measure the Hubble constant}

\author{Michael W. Coughlin}
\affiliation{School of Physics and Astronomy, University of Minnesota, Minneapolis, Minnesota 55455, USA}
\affiliation{California Institute of Technology, 1200 East California Blvd, MC 249-17, Pasadena, CA 91125, USA}

\author{Tim Dietrich}
\affiliation{Nikhef, Science Park, 1098 XG Amsterdam, The Netherlands}

\author{Jack Heinzel}
\affiliation{Artemis, Universit\'e C\^ote d'Azur, Observatoire C\^ote d'Azur, CNRS, CS 34229, F-06304 Nice Cedex 4, France}
\affiliation{Carleton College, Northfield, MN 55057, USA}

\author{Nandita Khetan}
\affiliation{Gran Sasso Science Institute (GSSI), I-67100 L'Aquila, Italy}

\author{Sarah Antier}
\affiliation{APC, UMR 7164, 10 rue Alice Domon et Léonie Duquet, 75205 Paris, France}

\author{Mattia Bulla}
\affiliation{Nordita, KTH Royal Institute of Technology and Stockholm University, Roslagstullsbacken 23, SE-106 91 Stockholm, Sweden}

\author{Nelson Christensen}
\affiliation{Artemis, Universit\'e C\^ote d'Azur, Observatoire C\^ote d'Azur, CNRS, CS 34229, F-06304 Nice Cedex 4, France}
\affiliation{Carleton College, Northfield, MN 55057, USA}

\author{David A. Coulter}
\affiliation{Department of Astronomy and Astrophysics, University of California, Santa Cruz, CA 95064, USA}

\author{Ryan J. Foley}
\affiliation{Department of Astronomy and Astrophysics, University of California, Santa Cruz, CA 95064, USA}

\begin{abstract}
The detection of GW170817 is revolutionizing many areas of astrophysics with the joint observation of gravitational waves and electromagnetic emissions. 
These multi-messenger events provide a new approach to determine the Hubble constant, thus, they are a promising candidate for mitigating the tension between measurements of Type Ia supernovae via the local distance ladder and the Cosmic Microwave Background. 
In addition to the ``standard siren'' provided by the gravitational-wave measurement, the kilonova itself has characteristics that allow to improve existing measurements or to perform yet another, independent measurement of the Hubble constant without gravitational-wave information. 
Here, we employ standardization techniques borrowed from the type-Ia community and apply them to kilonovae, not using any information from the gravitational-wave signal.
We use two versions of this technique, one derived from direct observables measured from the lightcurve, and the other based on inferred ejecta parameters, e.g., mass, velocity, and composition, for two different models.
These lead to Hubble Constant measurements of $H_0 = 109^{+49}_{-35}$\,km $\mathrm{s}^{-1}$ $\mathrm{Mpc}^{-1}$ for the measured analysis, and $H_0 = 85^{+22}_{-17}$\,km $\mathrm{s}^{-1}$ $\mathrm{Mpc}^{-1}$ and $H_0 = 79^{+23}_{-15}$\,km $\mathrm{s}^{-1}$ $\mathrm{Mpc}^{-1}$ for the inferred analyses.
This measurement has error bars within ~$\sim$\,2 to the gravitational-wave measurements ($H_0=74^{+16}_{-8}$\,km $\mathrm{s}^{-1}$ $\mathrm{Mpc}^{-1}$), showing its promise as an independent constraint on $H_0$.

\end{abstract}


\maketitle

A precise knowledge of the Hubble constant ($H_0$), to determine the expansion rate of the Universe, is one of the most important measurements driving the study of cosmology \cite{RiFi1998,PeAl1999}.
It has been known for a long time that the combined detection of gravitational waves (GWs) and their potential electromagnetic counterparts are useful for measuring the expansion rate of the universe \cite{Sch1986}.
These measurements are interesting since the GW standard siren measurements of $H_0$ do not rely on a cosmic distance ladder and do not assume any cosmological model as a prior.

This measurement has been made possible by the detection of GW170817~\cite{AbEA2017b} and AT2017gfo, a ``kilonova,'' which is thermal emission produced by the radioactive decay of neutron-rich matter synthesized from the ejecta of the compact binary coalescence at optical, near-infrared, and ultraviolet wavelengths \cite{2017ApJ...848L..21A,ChBe2017,CoBe2017,2017Sci...358.1570D,2017Sci...358.1565E,2017ApJ...848L..25H,2017Sci...358.1579H,KaNa2017,KiFo2017,2017ApJ...848L..20M,2017ApJ...848L..32M,NiBe2017,2017Sci...358.1574S,2017Natur.551...67P,SmCh2017,2017Natur.551...71T,2017PASJ...69..101U}. 
The analysis of GW170817 and the redshift of its host galaxy led to a measurement of $H_0=74^{+16}_{-8}$ km/s/Mpc (median and symmetric 68\% credible interval), 
where degeneracy in the GW signal between the source distance and the weakly constrained angle of inclination between the total angular momentum of the binary and the line of sight dominated the $H_0$ measurement uncertainty \cite{2017Natur.551...85A}. 
It has been estimated that $\sim 50$--$100$ GW events with identified optical counterparts would be required to have a $H_0$ precision measurement of $\sim 2\%$ \cite{ChFa2017}. 
Of course, these searches are a significant observational challenge because one has to cover, over 
a short time interval, a large localization region, typically larger than the $\sim\,20$ square degrees for GW170817~\cite{CoAh2019,CoAh2019b}.
The resulting $H_0$ measurements can be improved with, for example, high angular resolution imaging of the radio counterpart. Hotokezaka et al.~\cite{HoNa2018} applied this technique for GW170817 and obtained $H_0=68.9^{+4.7}_{-4.6}$ km/s/Mpc.

\begin{figure*}[t] 
 \includegraphics[width=7.0in]{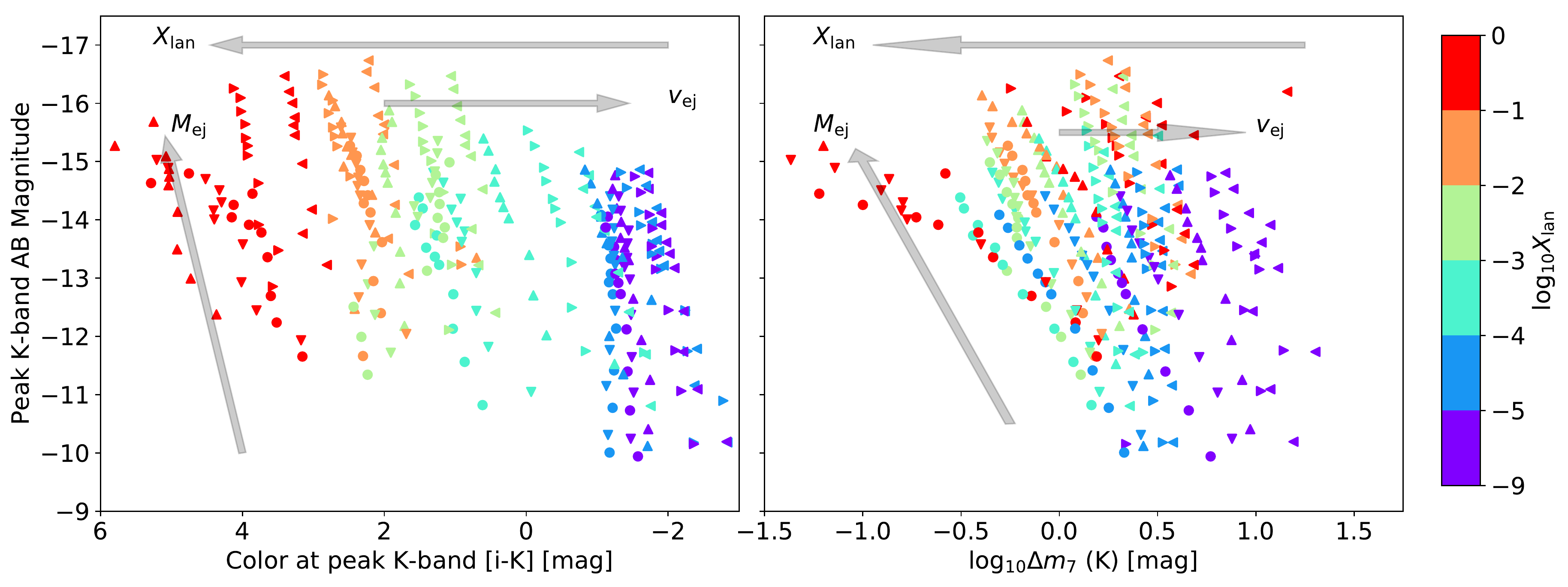}
  \caption{On the left is the color-magnitude diagram for the models in \cite{KaMe2017}. On the right is $\Delta m_{7}$ (measured in K-band over 7 days) - magnitude diagram for the same. Lanthanide fraction values $X_{\rm lan}$ = [$10^{-9}$,  $10^{-5}$, $10^{-4}$, $10^{-3}$, $10^{-2}$, $10^{-1}$] are shown by the color bar, while for $v_{\rm ej}$ = [0.03, 0.05, 0.1, 0.2, 0.3], the markers are circle, triangle, upside-down triangle, triangle pointing to the right, and triangle pointing to the left.}
 \label{fig:color_magnitude}
\end{figure*}

In this letter, we employ techniques borrowed from the type-Ia supernova community to measure distance moduli based on kilonova lightcurves. 
When constraining distances with type-Ia supernovae, it is required to anchor them to some other ``primary'' distance indicators in order to calibrate their absolute magnitude. 
In this work, we are using modelled light curves and here, the absolute magnitude is given by these models, and so we do not require to know their distances apriori by some other technique.
The method relies on differences in the modeled lightcurves due to ejecta parameters such as the ejecta mass ($M_\mathrm{ej}$), ejecta velocity ($v_\mathrm{ej}$), and lanthanide fraction ($X_\mathrm{lan}$).
Recently, it was also shown that a similar approach can be used for the bolometric luminosity to standardize kilonovae \cite{KaRa2019}.
Qualitatively, the imprint of the ejecta properties can be modeled by the semi-analytic methods of Arnett \cite{Arn1982}, and more broadly by these models' success at predicting the lightcurves for GW170817 (e.g. Ref.~\cite{SmCh2017}). Arnett-based models for kilonovae require a power term, which has generally been taken to be $P \propto t^\beta$ as appropriate for radioactivity models \cite{Met2017}, in addition to having other parameters such as the ejecta mass, the energy (or equivalently the velocity), and the opacity. Most important to  the overall luminosity is the heating rate per mass of the ejecta, determined by the product of intrinsic decay power and thermalization efficiency. Estimates of the thermalization efficiency based on simulations exist \cite{BaKa2013}, although they still are the largest systematic error budget, as the mass scales roughly inversely with the powering level. From Ref.~\cite{Arn1982}, the diffusion timescale is $\tau \propto \left(\frac{\kappa M}{v}\right)^{1/2}$, and similarly, the density is $\rho \propto \frac{M}{V} \propto \frac{M}{(v \tau)^3}$; in this way, all of these quantities affect the observables. For the simplest model where all of the energy is injected at $t=0$, corresponding to the time of peak luminosity, then the luminosity as a function of time $L(t) \propto L_0 e^{-t / \tau}$, which implies that $\log(L(t)/L_0) \propto -t / \tau$. This argues that the change in magnitude will be proportional to $\tau$.

Based on this, we explore color-magnitude diagrams for kilonovae, with the idea that measuring time constants and colors may be useful for determining the underlying luminosities.
In the left-hand panel of Fig.~\ref{fig:color_magnitude}, we show these quantities for all of the spherical models made available in Ref.~\cite{KaMe2017} plotting $i$-band minus $K$-band~\footnote{https://github.com/dnkasen/Kasen\_Kilonova\_Models\_2017}.
A few trends stand out:
As expected, the simulations with lower $X_{\rm lan}$ have lower absolute $K$-band magnitudes than those with higher $X_{\rm lan}$, with 1-2\,mag differences seen depending on the lanthanide fraction.
The much larger effect is on the color.
From the lowest to highest $X_{\rm lan}$, the $i$-band minus $K$-band color can vary by up to 5-6\,mag.
In addition, the velocities predominantly change the color, but at a much lower level, changing the color $\lesssim$\,1\,mag.
The trend with $M_{\rm ej}$ is a clear increase in peak magnitude, which is true of all lanthanide fraction and ejecta velocity pairs.
Moreover, the overall K-band peak luminosity increases as $X_{\rm lan}$ increases (or as one looks to the right in the grid). There is a similar but smaller trend with velocity.

In the right hand panel, we plot the change in luminosity, $\Delta m_{7}$, between peak and 7 days later in $K$-band.
As $X_{\rm lan}$ decreases, the effects on $\Delta m_{7}$ increase, with $\Delta m_{7} \gtrsim 1$\,mag at low $X_{\rm lan}$ and $\Delta m_{7} \lesssim 1$\,mag at higher $X_{\rm lan}$.
We plot the peak K-band magnitudes vs.\ ejecta mass for the available lanthanide fractions and ejecta velocities of the employed simulation set in the Supplementary material (this is essentially the same plot as Fig.~\ref{fig:color_magnitude}, but with the points separated out by ejecta velocity and lanthanide fraction).

The clear linear structure in Fig.~\ref{fig:color_magnitude} motivates the potential for their standardization, similar to SNe Ia lightcurves  (see e.g. Ref.~\cite{BuPa2018}). 
In the case of SN Ia lightcurves, they typically reach peak 17 days after explosion and then decay on a timescale of a few months. 
This motivates the use of the peak brightness, the time of the peak, and the ``width'' of the lightcurve as characteristic variables that can be compared, as realized early on \cite{RiFi1998,PeAl1999}. 
Similar to the decline-rate parameter used in the SN 1a community (typically $\Delta m_{15}$), we will define a K-band decay parameter over 7 days as discussed above.
This has the benefit of being measured from the observed light curve, with a downside of being tied to a particular filter and photometric system.
It also requires the peak in this passband to be well-measured, which is perhaps more straight-forward in the near-infrared where the peaks occur after a few days and therefore may be identified more easily. 
One downside might be that this band is less likely to be imaged in typical follow-up observations (i.e., before a kilonova has been confidently identified) because of the lack of infrared imagers on typical telescopes.
Therefore, while our analysis makes one choice, there could be others more suitable 
for the particular observational situation.

For the moment, we will ignore the so-called K corrections that arise from the fact that the observed spectral energy distribution is redshifted by a factor (1 + z) and are effectively observing with filters that have been blueshifted in the rest frame of the kilonova. Given the local sensitivity volume for these kilonovae, this is reasonable, although their inclusion could be straight-forward by adopting a spectral energy distribution.
A similar concern is that we implicitly ignore dust reddening in the analysis.
In general, the changes in color due to the evolution of the kilonova is much faster than the evolution of the dust reddening timescales.
This color term can be disentangled by reasonably high-cadence, multi-color imaging to measure the dust component that follows a typical reddening law and is constant in time.

\begin{figure}[t] 
 \includegraphics[width=3.6in]{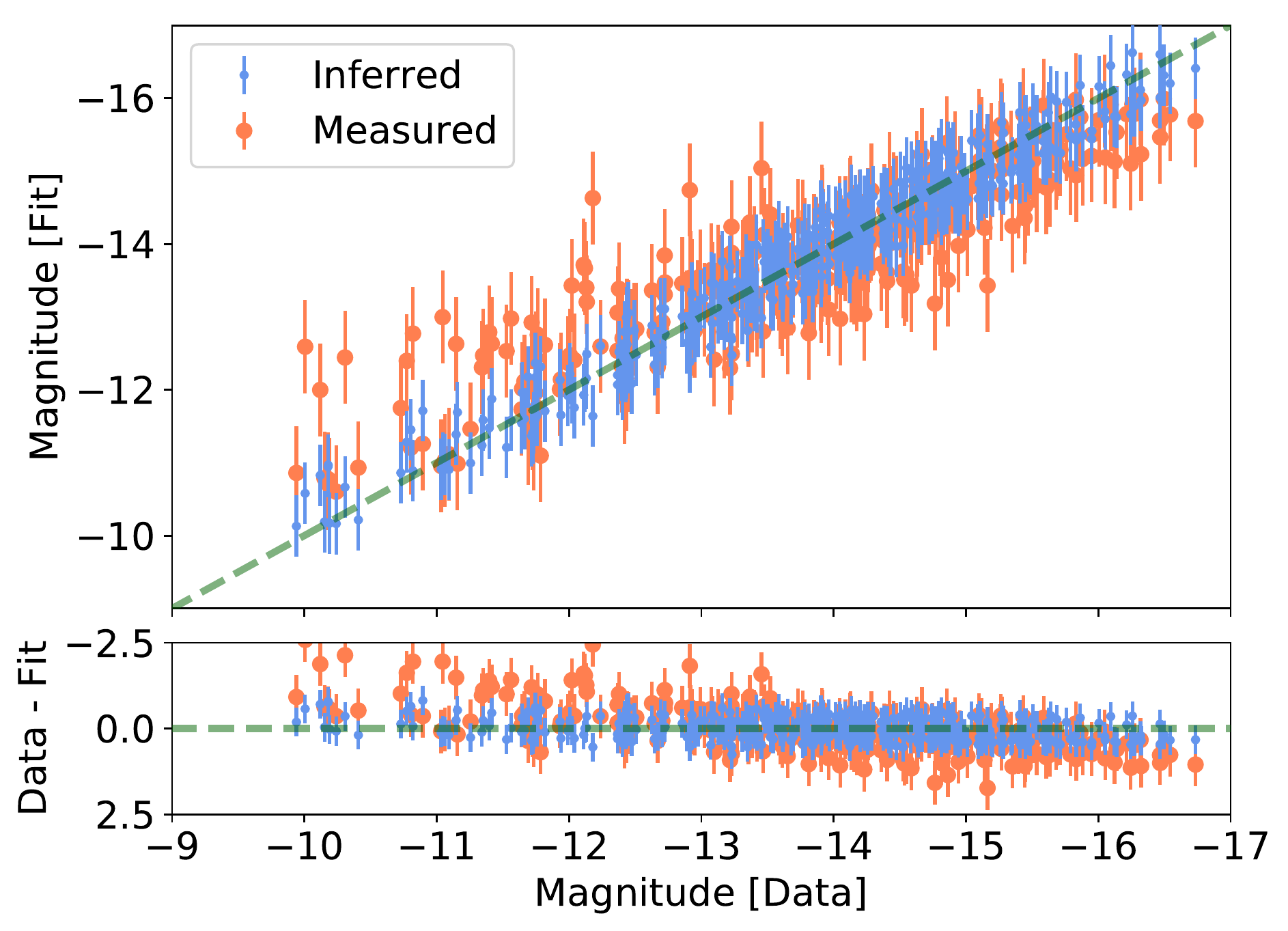}
  \caption{The top panel shows the ``Fundamental plane'' plot for the fit of Eqs.~\ref{eq:fit_measured} and ~\ref{eq:fit_inferred} and to the color-magnitude diagram shown in Figure~\ref{fig:color_magnitude} for the models in \cite{KaMe2017}. The ``measured''  values are derived from direct observables measured from the lightcurve, and the ``inferred'' values are from model-dependent ejecta parameters (mass, velocity, and lanthanide fraction). The bottom panel shows the difference between the computed values from the model and the fits for the simulations analyzed here. 
  }
 \label{fig:fundamental}
\end{figure}

We adopt two models where we do regressions in three dimensions in order to fit a distance for the kilonovae.
The first is based purely on observable quantities, while the second will be on quantities inferred from the model-based lightcurve fitting.
We note that these points in parameter space are given equal weights in the model, and therefore implicitly assume that all portions of the parameter space are equally likely.
To do this, we use a Gaussian Process Regression (GPR) based interpolation \citep{DoFa2017} to create a fit to the peak K-band magnitude for arbitrary parameters. 
This takes the form:
\begin{equation}
\begin{aligned}
M_{\mathrm{K}=\mathrm{K}_\mathrm{max}} = & f(\log_{10} \Delta m_{\mathrm{K}=7}, \log_{10} \Delta m_{\mathrm{i}=7}, \\
& m_{\mathrm{i}=\mathrm{K}_\mathrm{max}} - m_{\mathrm{K}=\mathrm{K}_\mathrm{max}})
\end{aligned}
\label{eq:fit_measured}
\end{equation}
We employ a GPR based interpolation instead of a linear fit due to the significant covariance between parameters.
As discussed above, the quantities we fit are $\Delta m_{7}$ in K-band, $\Delta m_{7}$ in i-band, and the $i$ minus $K$-band color at peak in K-band.
In order to compare with the apparent magnitudes, and therefore compute a distance modulus
\begin{equation}
M_{\mathrm{abs}} = m_{\mathrm{app}} - \mu
\label{eq:modulus}
\end{equation}
where $m_{\mathrm{app}}$ is the apparent magnitude and $\mu = 5 \log_{10} (\frac{D}{10 \mathrm{pc}})$ is the desired distance modulus (here, $D$ is the distance).
In principle, after the $\Delta m_{7}$ and color-dependent corrections have been applied, there will remain an intrinsic dispersion in the lightcurves, perhaps arising from location in the galaxy or perhaps dependent on the galaxy type.

We compare this ``measured'' fit to a fit based on ``inferred'' quantities of ejecta mass, ejecta velocity, and lanthanide fraction:
\begin{equation}
M_{\mathrm{K}=\mathrm{K}_\mathrm{max}} = f(\log_{10} M_{\rm ej},v_{\rm ej},\log_{10} X_{\rm lan})
\label{eq:fit_inferred}
\end{equation}
We note that this is not specific to these observables; for example, Ref.~\cite{Bul2019}\footnote{https://github.com/mbulla/kilonova\_models} use ejecta mass $M_{\rm ej}$, the temperature at 1~day after the merger $T_0$~\footnote{In this analysis, we use simulations from \cite{Bul2019} where the temperature is taken as a parameter rather than calculated from the adopted densities and nuclear heating rates. We will address this point and use self-consistent models in future analyses.}, the half-opening angle of the lanthanide-rich component $\Phi$ (with $\Phi=0$ and $\Phi=90^\circ$ corresponding to one-component lanthanide-free and lanthanide-rich models, respectively) and the observer viewing angle $\theta_{\rm obs}$ (with $\cos\theta_{\rm obs}=0$ and $\cos\theta_{\rm obs}=1$ corresponding to a system viewed edge-on and face-on, respectively). We will compare these two models in the following.
For consistency, we also employ a GPR based interpolation here, although we have found that a linear fit based on these variables gives comparable results.
When performing the analysis, we include an overall error during the fit that represents scatter from intrinsic variability in the kilonovae models. These errors are $\sim$\,0.7\,mag for the ``measured'' case and $\sim$\,0.4\,mag for the inferred case, which are derived from the median error reported by the GPR across the parameter space.
Fig.~\ref{fig:fundamental} shows the performance of the fits of Eqs.~\eqref{eq:fit_measured} and ~\eqref{eq:fit_inferred} compared to the models in \cite{KaMe2017}.
In general, there is some systematic scatter in the ``measured'' case, indicating that the $\Delta m_{7}$ in K-band, $\Delta m_{7}$ in i-band, and the $i$ minus $K$-band color at peak in K-band are not sufficient to completely standardize across the parameter space. In all likelihood, the proper choice of variables for this purpose will become more apparent as detections are made.
The fits broken up by lanthanide fraction and velocity for the inferred and measured cases are shown in the Supplementary material.

\begin{figure}[t] 
 \includegraphics[width=3.5in]{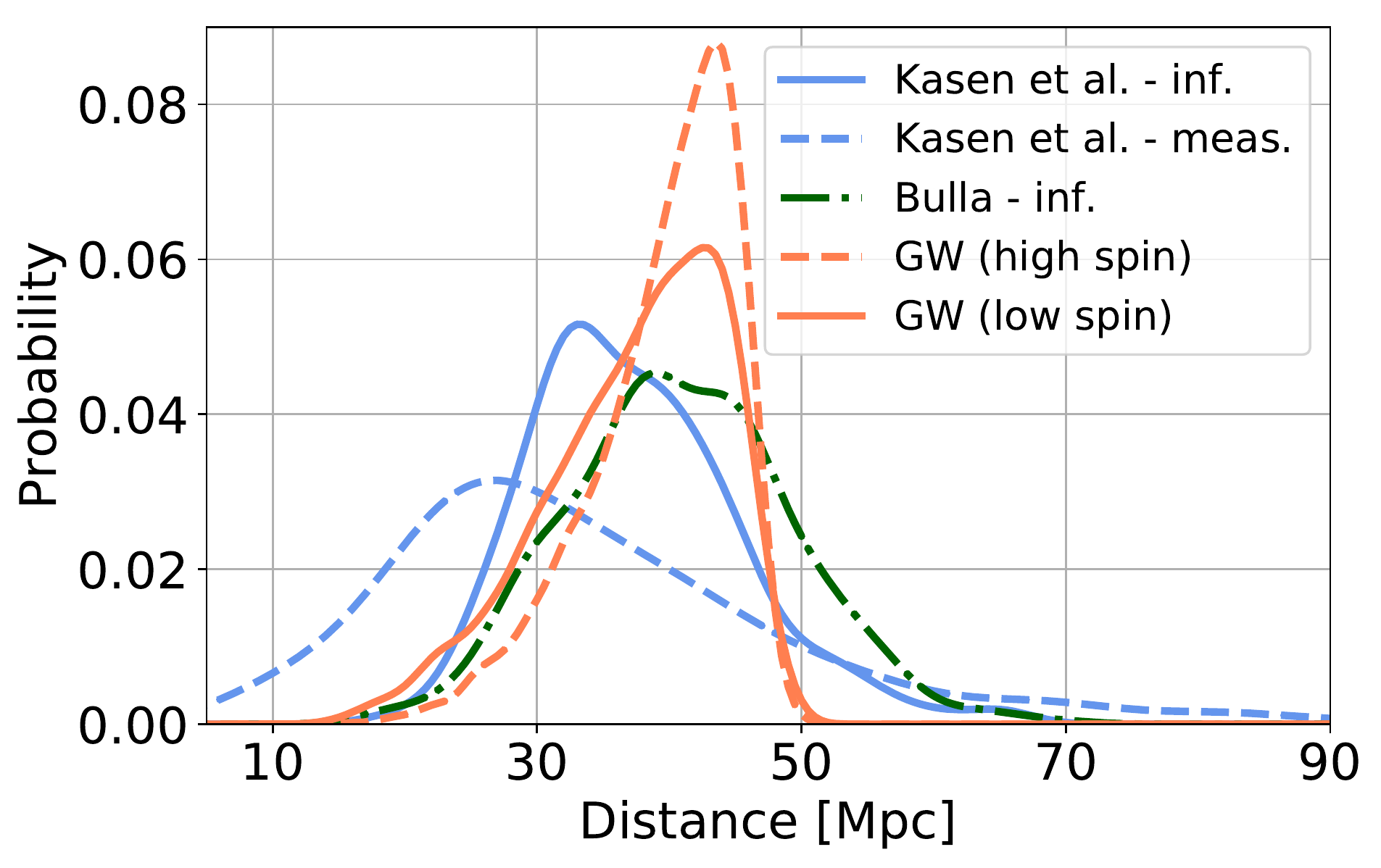}
  \caption{Posterior distributions for distance to GW170817, where the results of the GW-only analyses (high spin, dashed and low spin, solid) \cite{AbEA2018}, the Kasen et al. \cite{KaMe2017} kilonova-only analyses (measured, dashed and inferred, solid), and the Bulla \cite{Bul2019} kilonova analysis based on the fit of Eq.~\ref{eq:fit_inferred} are shown. 
  }
 \label{fig:distance}
\end{figure}

We use the fit of Eqs.~\eqref{eq:fit_measured} and ~\eqref{eq:fit_inferred} and apply them to the posteriors on ejecta mass, velocity, and lanthanide fraction derived previously~\cite{CoDi2017,CoDi2018,CoDi2018b}. 
In these fits, we assume a 1.0\,mag systematic uncertainty on the model; in this sense, we do not require that the models are ``perfect'' but instead encode some of the systematic uncertainties associated with them.
We sample from the posteriors, in addition to the distributions for the measured parameters, to derive a constraint of $D = 31^{+17}_{-11}$\,Mpc for the Kasen et al. \cite{KaMe2017} measured analysis. We find $D = 37^{+8}_{-7}$\,Mpc and $D = 40^{+9}_{-8}$\,Mpc for the Kasen et al. \cite{KaMe2017} and Bulla \cite{Bul2019} inferred analyses respectively.
This is consistent (given the relatively broad error bars) with other measurements of the host galaxy for GW170817, e.g. \cite{HjLe2017,LeLy2017,CaJe2018}.
We show this constraint in Fig.~\ref{fig:distance} along with the high and low spin posteriors presented in \cite{AbEA2018}.

\begin{figure}[t] 
 \includegraphics[width=3.5in]{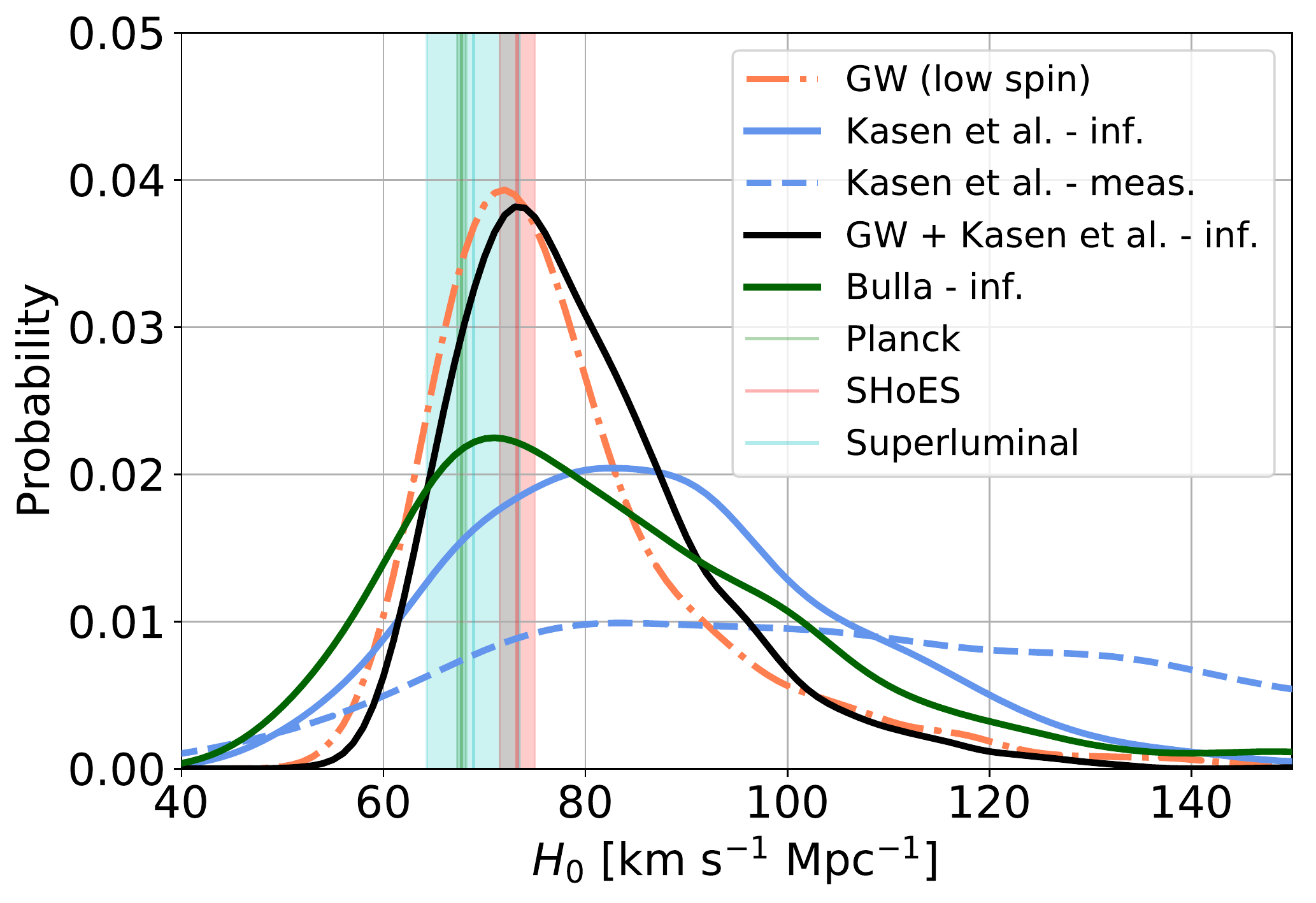}
  \caption{Posterior distributions for $H_0$ for GW170817, where the results of the GW-only analysis, the Kasen et al. \cite{KaMe2017} and Bulla \cite{Bul2019} models kilonova-only analyses, and the Kasen et al. \cite{KaMe2017} combined GW-EM analysis are shown. The 1- and 2-$\sigma$ regions determined by the ``superluminal'' motion measurement from the radio counterpart (blue) \cite{HoNa2018}, Planck CMB (TT,TE,EE+lowP+lensing) (green) \cite{Ade2015} and SHoES Cepheid-SN distance ladder surveys (orange) \cite{RiMa2016} are also depicted as vertical bands.
  }
 \label{fig:H0}
\end{figure}

Following the analysis of \cite{2017Natur.551...85A}, we compute the corresponding values of the local Hubble constant for the kilonova analyses.
For the measured analysis, we show the Kasen et al. \cite{KaMe2017} kilonova-only Hubble constant measurement of $H_0 = 109^{+49}_{-35}$\,km $\mathrm{s}^{-1}$ $\mathrm{Mpc}^{-1}$, median and symmetric 68\% credible interval, in Fig.~\ref{fig:H0}.
For the inferred analyses, we also show the Kasen et al. \cite{KaMe2017} and Bulla \cite{Bul2019} analyses, giving $H_0 = 85^{+22}_{-17}$\,km $\mathrm{s}^{-1}$ $\mathrm{Mpc}^{-1}$ and $H_0 = 79^{+23}_{-15}$\,km $\mathrm{s}^{-1}$ $\mathrm{Mpc}^{-1}$. Models from Kasen et al. \cite{KaMe2017} and Bulla et al. \cite{Bul2019} differ in several aspects, with the latter assuming parameterized opacities and temperatures, adopting different density profiles and ejecta geometries and taking into account the interplay between the two ejecta components. While the use of two different kilonovae models is not a truly independent analysis, the fact that they are different but the corresponding distributions consistent with one another gives confidence that the analysis is not particularly model dependent.

We also compute the corresponding values of the local Hubble constant for an analysis which combines the kilonova and GW derived distances (using the high-spin posteriors, as done in \cite{2017Natur.551...85A}).
Because the GW and kilonova data are independent, the posterior probabilities for the distances can simply be multiplied.
For the Kasen et al. \cite{KaMe2017} inferred analysis, we find a combined measurement of $H_0 = 78^{+14}_{-9}$\,km $\mathrm{s}^{-1}$ $\mathrm{Mpc}^{-1}$,
while for the measured analysis, the results are the same as for the GW analysis.
Altogether, this proves that the kilonova measurement is competitive with the GW measurements 
to obtain an independent constraint on $H_0$.

In this article, we have demonstrated how to use parameters derived from radiative transfer simulations to give distance measurements using kilonovae, see also~\cite{KaRa2019} for an alternative approach. We have adopted the peak luminosity in K-band, the decay in K-band over 7 days, $\Delta m_{7}$, and the $i$ minus $K$-band colors.
We have shown that these distance estimates are consistent with other measurements of GW170817's host galaxy directly and provide competitive measurements of $H_0$ to GW distance measurements alone.

These techniques will play an important role in an era of both large-scale optical surveys, and radio-to-X-ray follow-up efforts, in identifying electromagnetic counterparts of compact binary coalescences. 
The Large Synoptic Survey Telescope (LSST) \cite{Ivezic2014} and the Wide-Field InfraRed Survey Telescope (WFIRST) \cite{SpGe2015} in particular would be able to identify candidates with optical/NIR emission similar to GW170817 beyond 300\,Mpc (i.e., the limit of current Advanced LIGO GW surveys for compact binary coalescences). 
Techniques such as those described here will play a key role in making these detections useful for studies of the Hubble constant and, thus, Dark Energy.

To explore this further, we examine the potential constraints over a sample of similar events.
First of all, we assume that the 26\% measurement here is representative of the typical width of the $H_0$ measurement from an individual event; this might be optimistic as AT2017gfo has a very well-sampled lightcurve.
We also assume that the constraints converge as $\frac{1}{\sqrt{ \mathrm{N}}}$, where N is the number of kilonovae.
Ref.~\cite{ChFa2017} note that for the GW counterpart measurement, the typical tails in the distributions ``average out'' to make a smaller effective width than is usual for any given single event, so this may be a conservative assumption.
Under these assumptions, the number of events required to make a 6\% and 2\% measurement is $\sim$\,20 and $\sim$\,170 respectively.
For the sake of comparison, Ref.~\cite{ChFa2017} found that 50-100 binary neutron stars with counterparts are required to make a 2\% measurement.
They also found that for binary neutron stars without counterparts, where a galaxy-catalog based statistical method is required, $\sim$\,50 binary neutron stars are required to make a 6\% measurement.
In this sense, it is less powerful than binary neutron stars with counterparts but more so than binary neutron stars without counterparts.
It is also likely that in practice, the error bars on a kilonova only analysis will become smaller with time as the models improve.
Further detections of kilonovae will also make it possible to perform fits for Eq.~\eqref{eq:fit_measured} directly using data rather than using models, which may also yield improved constraints.

There are other techniques that could be employed that may yield further constraints.
Ref.~\cite{KaRa2019} uses a combination of semi-analytic kilonovae models with ejecta mass and velocity fits to numerical relativity simulations to show correlations between the maximum bolometric luminosity and decline from peak luminosity for a simulated population of binary neutron star mergers, showing that it may be possible to incorporate either models or future observations into the standardization of bolometric lightcurves.
In addition, Ref.~\cite{Doc2019} laid the groundwork for techniques for performing joint standard candle-standard siren measurements that bring together the analyses of GWs and their electromagnetic counterparts, demonstrating in particular their importance in case where the selection effects from kilonova and GW observations can be significant.

In addition, these techniques can be used to augment GW detector calibration, which are currently limited to a few percent in amplitude and a few degrees in phase in the latest observing runs and are reaching their fundamental limits in performance \cite{CaBe2017,AcEA2018}.
The match between the waveform predicted by general relativity and the GW strain signal alone can calibrate a single detector's relative amplitude and phase responses to a GW as a function of frequency \cite{EsHo2019}.
This GW strain signal unaccompanied by other messenger signals can also calibrate relative responses between two GW detectors.
However, when augmented by independent measurements of the event's distance and inclination angle, the detectors' absolute amplitude and phase responses to GWs can be calculated.
Therefore, our techniques could be a critical method in the effort to calibrate responses of GW detectors using astrophysical signals.
Although there will need to be many more GW detections with electromagnetic counterparts to improve the precision of the astrophysical calibration measurements over the current existing \textit{in situ} measurements, single events can also be used to improve and verify \textit{in situ} measurements.
Ref.~\cite{EsHo2019} showed that GW data alone can constrain the relative amplitude calibration uncertainty to less than 10\% with 10-20 events (less than 1\% with 1000-2000 events); when combining conservative constraints with broad distance and inclination constraints from electromagnetic data, it takes 400-600 events. Given the similar distance constraints assumed to those derived here, this is the right ballpark for this level of calibration uncertainty.

Further work is needed to understand how our restriction to a spherical geometry with a single-component would change for multiple components including possibly reprocessing between ejecta components, e.g.~\cite{KaSh2018}.
Discovering more kilonovae will improve the understanding of the effect of ejecta geometries, supporting the use of multiple components and informing how they should be included.
In addition, the standardization assumes that the radiative-transfer models are consistent to produce proper absolute magnitudes and colors (at least within the assumed error bars), which motivates continued work to improve the accuracy of the models and the grids that they are simulated on.
The potential for more sources, at the price of higher systematics, with this method leads to a trade-off between the use of kilonovae or purely GW measurements; GW measurements will have relatively lower levels of systematics at the price of fewer objects to use.
This method can be used in particular for any kilonovae detection with measurements of the host galaxy redshift, by way of fits of the ejecta parameters (e.g. Ref.~\cite{AsCo2018}).
It may also be possible to constrain the inclination using associated GRB detections or upper-limits on potential sub-threshold gamma-ray transients.
As both of these analyses will be directly testable by future kilonovae observations, the utility of analyses of this type will be dependent on the coming comparisons.

\emph{Acknowledgements.} 
The authors would like to thank Daniel Kasen for making his models publicly available.
MC is supported by the David and Ellen Lee Postdoctoral Fellowship at the California Institute of Technology.
TD acknowledges support by the European Union's Horizon 2020 research and innovation program under grant agreement No 749145, BNSmergers. 
NC and JH acknowledge support from the National Science Foundation with grant number PHY-1806990. SA is supported by the CNES Postdoctoral Fellowship at Laboratoire Astroparticle et Cosmologie.
The UCSC team is supported in part by NASA grant NNG17PX03C, NSF grants AST-1518052 and AST-1911206, the Gordon \& Betty Moore Foundation, the Heising-Simons Foundation, and by a fellowship from the David and Lucile Packard Foundation to R.J.F.
DC acknowledges support from the National Science Foundation Graduate Research Fellowship under Grant DGE1339067.

\bibliographystyle{unsrt85}
\bibliography{references}

\clearpage
\appendix

\begin{figure*}[t] 
 \centering
 \includegraphics[width=4.5in]{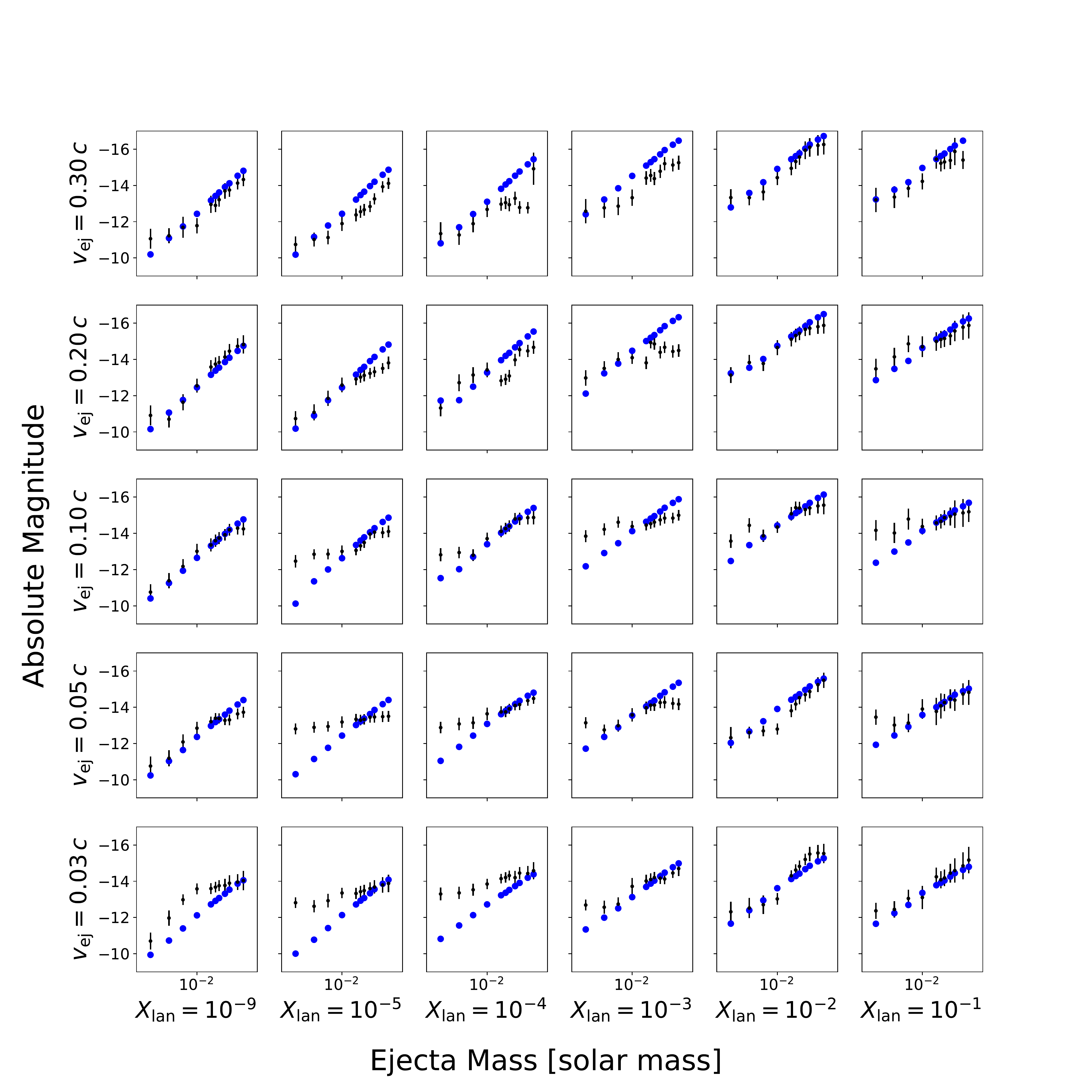}
 \includegraphics[width=4.5in]{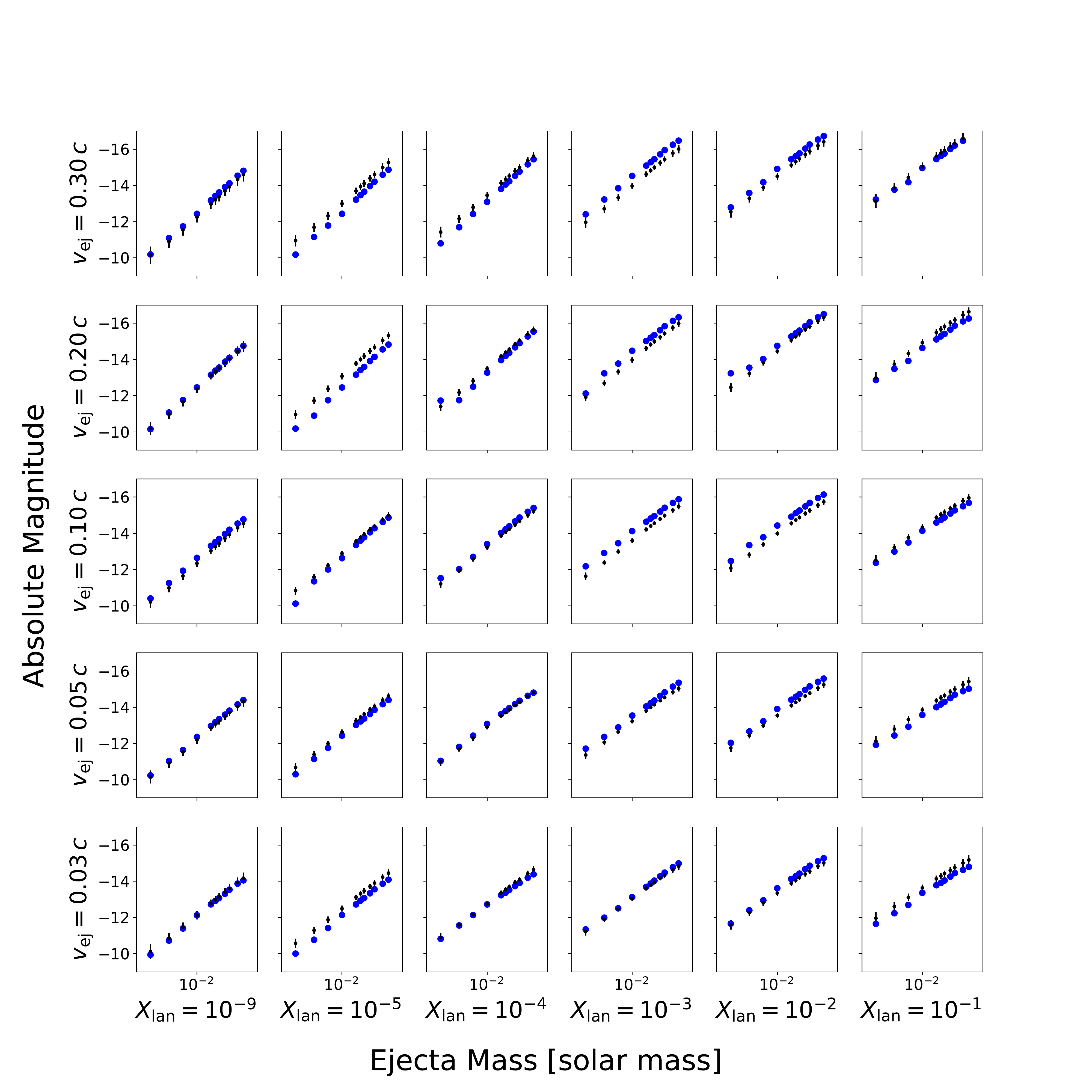}
  \caption{The fit of Eqs.~\eqref{eq:fit_measured} and ~\eqref{eq:fit_inferred} and to the color-magnitude diagram shown in Figure~\ref{fig:color_magnitude} for the models in Ref.~\cite{KaMe2017}, varying the available lanthanide fractions and ejecta velocities of the employed simulation set. The black points are the fits (with the measured error bar from the chi-squared) while the blue points are computed directly from the models in Ref.~\cite{KaMe2017}.}
 \label{fig:fits}
\end{figure*}

\end{document}